\documentstyle[12pt]{article}
\begin{document} 
\title{Monte Carlo Simulation of Ising Models with Dipole Interaction}
\author{U. Nowak and A. Hucht \\
Theoretische Tieftemperaturphysik, \\
Universit\"{a}t Duisburg,\\
47048 Duisburg / Germany\\
e-mail: uli@thp.uni-duisburg.de }
\date{}
\maketitle
\begin{center} {\large \bf Abstract} \end{center}
{\footnotesize
  Recently, a new memory effect was found in the metamagnetic domain
  structure of the diluted Ising antiferromagnet $Fe_x Mg_{1-x}
  Cl_2$ by domain imaging with Faraday contrast. Essential for this
  effect is the dipole interaction. We use a modified Monte Carlo
  algorithm to simulate the low temperature behavior of diluted
  Ising-antiferromagnets in an external magnetic field. The algorithm
  is specially adjusted to take care of long range interaction. The
  metamagnetic domain structure occurring due to the dipole
  interaction is investigated by graphical representation.

  In the model considered the antiferromagnetic state is stable for
  an external magnetic field smaller than a lower boundary $B_{c1}$
  while for fields larger than an upper boundary $B_{c2}$ the system
  is in the saturated paramagnetic phase, where the spins are
  ferromagnetically polarized.  For magnetic fields in between these
  two boundaries a mixed phase occurs consisting of ferromagnetic
  domains in an antiferromagnetic background. The position of these
  ferromagnetic domains is stored in the system: after a cycle in
  which the field is first removed and afterwards applied again the
  domains reappear at their original positions. The reason for this
  effect can be found in the frozen antiferromagnetic domain state
  which occurs after removing the field at those areas which have
  been ferromagnetic in the mixed phase.
}
\newpage

The threedimensional Ising model with an antiferromagnetic exchange
interaction undergoes a first-order phase transition from an
antiferromagnetic to a paramagnetic saturated phase for low
temperatures during an increase of the homogenous external magnetic
field. Because of demagnetizing field effects in experimental systems
like $FeCl_2$ there occurs a mixed phase for external fields $B_{c1}
\le B \le B_{c2}$ \cite{Dill}. In theoretical considerations the
existence of a mixed phase is often neglected since dipole
interactions have to be considered in order to investigate the mixed
phase. Especially in conventional Monte Carlo simulations the dipole
interaction can hardly be taken into account for lattices large enough
to investigate domain structures due to its long-range nature: for
each spin flip the number of operations to calculate the change in
energy scales with the number of spins in the system. Therefore we
developed a specially adjusted Monte Carlo algorithm which will be
published elsewhere to simulate spin-models with dipole interaction.

In this paper using this algorithm for the first time we perform
simulations in order to get a deeper understanding of a new memory
effect that was found recently in the mixed-phase domain structure of
the diluted Ising antiferromagnet $Fe_x Mg_{1-x} Cl_2$ by domain
imaging with Faraday contrast \cite{Matt}. The position and shape of
paramagnetic saturated domains which grow within the antiferromagnetic
state while field increasing is stored in the sence that the domains
reappear even after a cycle in which the field is first removed and
afterwards again applied. Essential for an investigation of this
effect is obviously the dipole interaction, since it is this long
range interaction which is responsible for the occurence of a mixed
phase.

The diluted Ising antiferromagnet in an external magnetic field (DAFF)
is an ideal system to study random field behavior theoretically as
well as experimentally since it is believed to be in the same
universality class as the random field Ising model (RFIM)
\cite{fish}.  A well known feature of the DAFF is the formation of a
domain state with extremely long relaxation times (for an overview see
ref. \cite{Bela}).  This domain state is frozen even for zero-field
and it is obtained by either cooling the system in an external field
from the paramagnetic high temperature phase or by decreasing the
field correspondingly. The mechanisms which are responsible for the
hysteretic properties of the DAFF have been investigated
experimentally \cite{birg,leit}, theoretically \cite{vill} and in
computer simulations \cite{gres,nowa1,nowa2,nowa3}. In the following
we will show that the understanding of the hysteretic behavior of the
DAFF is essential for an understanding of the memory effect.

The hamiltonian of an Ising model with dipole interaction in units of
the coupling constant reads

\begin{displaymath}
  H = \sum_{\langle i,j \rangle} \epsilon_{i} \epsilon_{j}
  \sigma_{i} \sigma_{j} - B \sum_{i} \epsilon_{i} \sigma_{i}
  + d \sum_{i,j} \frac{\epsilon_{i} \epsilon_{j}
  \sigma_{i}\sigma_{j}}{r_{i,j}^3} (1-3 \cos^2 \theta_{i,j})
\end{displaymath}

where $\sigma_{i}=\pm 1$ are the spins and the $\epsilon_{i} = 0,1$
represent the dilution $p = 5\%$. In the first sum $<i,j>$ means all
combinations of spins which are nearest neighbors. The exchange
interaction favoring antiferromagnetic alignment of spins is set equal
to one. The second sum represents the interaction with the external
magnetic field $B$.  The third sum is over all combinations of spins
and $d$ represents the strength of the dipole interaction ($d = 0.5$ in
this case). $r_{i,j}$ is the distance between two spins on sites $i$
and $j$ and $\theta$ is the angle between the $z$-axis (the direction
of the external field) and the distance vector $\underline{r}$. In
order to simplify the model we restrict ourselves to a two dimensional
system with open boundary conditions representing one plane of the
experimental system. As we will see the qualitative behavior of the
experimental system is well described by our model as far as the
domain structure is concerned, which is responsible for the memory
effect.

We use an antiferromagnetic long-range ordered system as the initial
spin configuration. The simulation is done at very low temperature,
$T=0.1$. The system builds up a saturated-domain state for a field of
$B = 2.65$ which is within the mixed phase. Then we investigate the
development of this domain state during a field cycle to zero field,
$B = 0$, and then back to the mixed phase, $B = 2.65$. The time that
is needed to equilibrate the system is a few hundreds of Monte Carlo
steps per spin.

The system has a size of $190 \times 50$ ($x \times z$), i.e 190
columns and 50 rows.  The figures show spin configurations of the
simulated system as well as the mean magnetisation of the columns of
the system. The external magnetic field is aligned with the easy axis
of the spins, the $z$-direction. Each site of the two dimensional
square lattice is represented by a square, the vacancies of the system
are shown as black, up-spins as grey and down-spins as white squares.

The mixed phase (Fig.1) consists of antiferromagnetic domains
(checkerboardlike) and paramagnetic saturated (ferromagnetic) domains
(grey, "spin-up"). In the latter domains the spins are aligned with
the field. The domains have the form of stripes . This follows
directly from the nature of the dipole interaction which favors those
spins to order ferromagnetically which are on lattice sites placed
along the direction of the easy axis leading to the development of
ferromagnetically ordered columns.  Also shown in the upper part of
Fig.1 is the column-magnetisation, i.e. the mean magnetisation of each
column of the system. This quantity corresponds to the Faraday
constrast that is observed in experiments. Since the domains are
striped there is a sharp contrast between antiferromagnetic domains
(magnetisation $\approx 0$) and paramagnetic saturated domains
(magnetisation $\approx 1/$per spin).

Lowering the field to zero the saturated domains vanish (Fig.2) and
the magnetisation decreases to nearly zero. However, an accurate
analysis of this zero-field spin configuration shows that it is not
completely antiferromagnetically long-range ordered. Instead, those
regions of the system, which have been paramagnetic saturated in the
field consist now of an antiferromagnetic domain structure
corresponding to the domain state of a DAFF after field decreasing. Why
does this happen ?

In a field-decreasing procedure the antiferromagnetic-para\-magnetic
phase boundary is crossed in a direction from the paramagnetic to the
antiferromagnetic phase. In this case, due to the unconventional
dynamics of the DAFF which follows from random-field pinning the
system cannot develop a long-range ordered state. Instead it freezes
in an antiferromagnetic domain state. This effect is the reason for
the unconventional structure of the system after removing the external
field.  In regions of the system which have been paramagnetic
saturated in the mixed phase a frozen antiferromagetic domain state
develops while in the regions of the system which have been
antiferromagnetic nothing changes, the long-range order persists. The
nonexponential dynamics of antiferromagnetic domains in a DAFF after
removing the external field has been investigated earlier (see
\cite{Han} and references therein). The domains are frozen and remain
practically constant on time scales accessible for observation. Note
that due to the existence of antiferromagnetic domains and domain
walls, respectively, there is a finite column-magnetisation. This
magnetisation is small compared to the magnetisation of a saturated
domain but it is larger than the magnetisation of an antiferromagnetic
column which is also not exactly zero due to fluctuations of the
vacancy distribution in our finite system.

After applying the magnetic field again a configuration of striped
antiferromagnetic domains arises once more (Fig.3).  Comparing this
configuration with the original domain configuration (Fig.1) one finds
that the original domain configuration is nearly reproduced, at least
in that sence that paramagnetic saturated domains grow first at those
places which have been paramagnetic saturated before.  This is the
memory effect. Its origin is the antiferromagnetic domain
configuration of the system after removing the field (Fig.2). The
regions which consist of an antiferromagnetic domain state are less
stable than the long range ordered regions since the first contain
domain walls. These regions are the first in which saturated domains
during an increase of the external field occur, restoring the original
metamagnetic domain configuration.

\section*{Acknowledgement}
This work was supported by the Deutsche Forschungsgemeinschaft through
Sonderforschungsbereich 166.

\newpage
{\footnotesize

\newpage
\section*{Figure Captions}

\begin{itemize} 
\item[FIG.1.] Spin configuration and column-magnetisation of the
  simulated system as explained in the text: a metamagnetic domain
  configuration in the mixed phase
\item[FIG.2.] Spin configuration and column-magnetisation of the
  simulated system: an antiferromagnetic domain configuration after
  removing the field
\item[FIG.3.] Spin configuration and column-magnetisation of the
  simulated system: the metamagnetic domain configuration after
  applying the field again
\end{itemize}
}
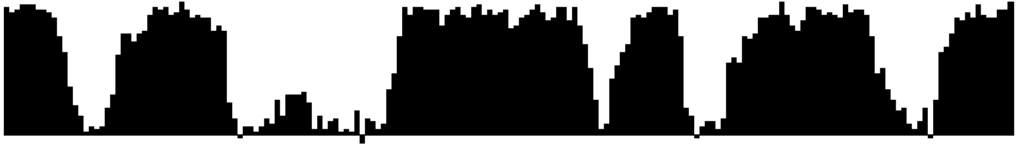
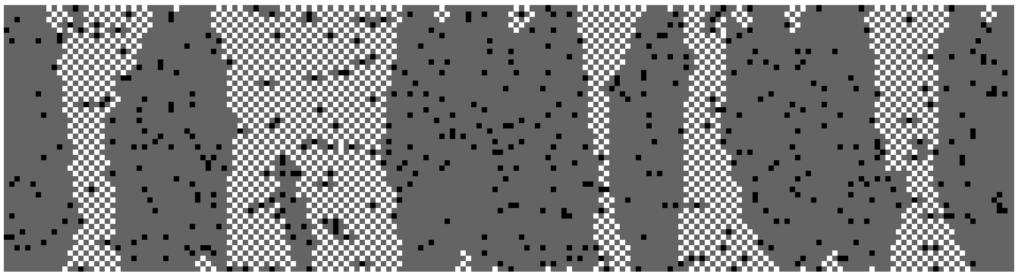
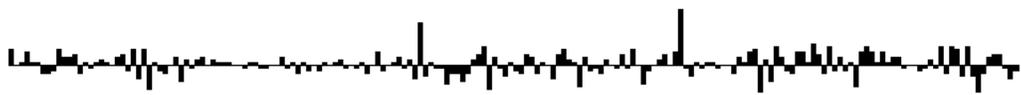
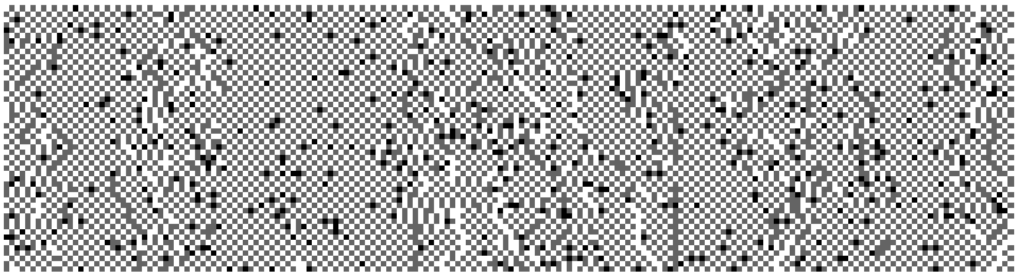
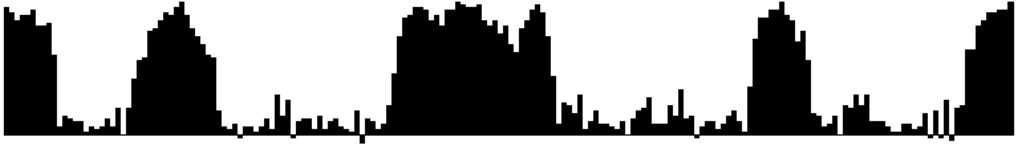
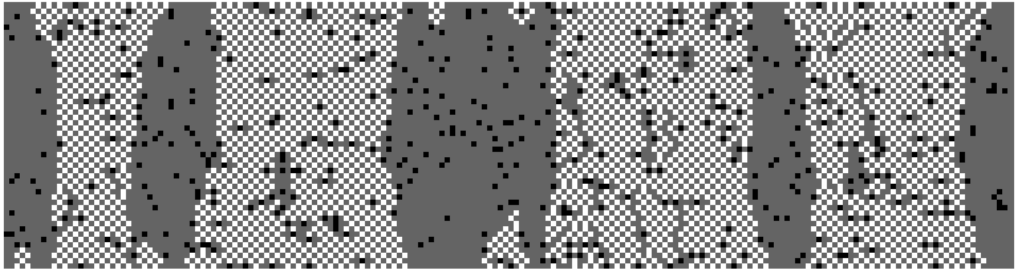
\end{document}